\documentclass[12pt,thmsa]{article}

\input{tcilatex}

\input tcilatex
\begin{document}

\title{Modular Theory and Geometry}
\author{B. Schroer and H.-W. Wiesbrock \\
FB Physik, FU Berlin, Arnimallee 14, Berlin 14195}
\date{June, 1998}
\maketitle

\begin{abstract}
In this letter we present some new results on modular theory and its
application in quantum field theory. In doing this we develop some new
proposals how to generalize concepts of geometrical action.

Therefore the spirit of this letter is more on a programmatic side with many
details remaining to be elaborated.
\end{abstract}

\section{Introduction}

The basis of this new structure in QFT is modular theory. Mathematically it
is a vast generalization of the modular factor which accounts for the
difference between right and left Haar measure in the case of non unimodular
groups as e.g. the group of upper triangle matrices. Tomita and later
Takesaki succeeded to convert this idea into a powerful tool for the
investigation of von Neumann algebras, see \cite{Ta}. In fact Alain Connes
could not have carried out his pathbreaking work on the classification of
factor algebras without this theory. Modular theory is also looming behind
the subfactor theory of Vaughn Jones. The physics side is equally
impressive. At the time when Tomita presented his theory, Haag Hugenholz and
Winnink published their fundamental work on (heat bath) thermal aspects of
QFT, see \cite{H/H/W}. The KMS condition (a name which they coined) was used
up to that time by various physicist (in particular Kubo, Martin and
Schwinger) as a clever mathematical trick in order to avoid to compute
cumbersome traces in evaluating Gibbs thermal ensembles. In the hands of
Haag Hugenholz and Winnink this formula became the key for their formulation
of equilibrium quantum statistical mechanics directly in the thermodynamic
limit for which the Gibbs representation becomes meaningless. The generator
for the modular operator turns out to be a thermal hamiltonian with
two-sided spectrum and finite fluctuations which acts on the original
algebra as well as on its commutant which represents a kind of shadow world
(which corresponds in the above analogy to group theory to the conjugate
action ) \cite{Haa}. (See \cite{Ja} for new results on the physical
interpretation of the underlying shadow world.) The Tomita antiunitary
involution J turns out to be the normalized flip operation which maps the
left into the right action. Later Bisognano and Wichmann \cite{Bi/Wi} found
another very nontrivial field theoretic illustration of the Tomita-Takesaki
theory in their study of the algebra which is generated by fields restricted
to the wedge region (i.e. the same (Rindler) region which played an
important role in Unruh's discussion of the Hawking effect). In their case
the modular group turned out to be the wedge-affiliated Lorentz boost and
the Tomita $J$ was (up to a 90 degree rotation around the boost direction)
the field theoretical TCP operator. Later Sewell showed that this beautiful
result of Bisoganano and Wichmann might be seen as a generalization of the
Unruh effect, see\cite{Se,Un}. (Below we present an algebraic criterion
which implies this property, see also \cite{Bo3}) This time the shadow-world
(the von Neumann commutant) was part of the real world namely the algebra of
the causally disjoint region behind the wedge horizon. Besides having a
thermal meaning the modular theory got a beautiful geometrical
interpretation. This work (as well as some special prior observations on
free fields \cite{O/L/R}) strongly suggested that there was a deep relation
between modular theory and relativistic causality and localization. In more
recent times Borchers \cite{Bo1} and one of the present authors (H-W. W.)%
\cite{Wi2} turned the geometrical action of modular theory around and showed
that with two subalgebras in appropriate modular position one can build up
spacetime symmetries (d=1+1 conformal \cite{G/L/W} and, with somewhat
stronger assumptions, also the higher dimensional cases \cite{Ka/Wi}\cite
{Wi4}). Starting from Wigner's representation theory of positive energy
representations it is possible to construct the unique net associated with
the (m,s=semiinteger) Wigner representation without using (nonunique) field
coordinates as the generators of local algebras \cite{Schr1}.

We believe that these deep relation between space-time symmetry and pure
quantum physical properties will be important for the understanding of the
still evasive ''Quantum Gravity''. The modular structure also promises to
clarify some points concerning the physics of the Wightman domain properties 
\cite{Schr1}. In fact the modular groups act linearity on the ''field
space'' i. e. the space generated by applying a local field on the vacuum.
Therefore this space, which is highly reducible under the Poincar\'{e}
group, may according to a conjecture of Fredenhagen (based on the results in 
\cite{Fr/Joe}) in fact carry an irreducible representation of the union of
all modular groups (an infinite dimensional group $\mathcal{G}_{mod}$ which
contains in particular all local spacetime symmetries). The equivalence of
fields with carriers of irreducible representations of an universal $%
\mathcal{G}_{mod}$ would add a significant conceptual element to LQP and
give the notion of quantum fields a deep role which goes much beyond that of
being simply generators of local algebras. Our arguments suggest that in
chiral conformal QFT $\mathcal{G}_{mod}$ includes all local diffeomorphism.
A related group theoretical approach this time starting from properties of
modular involutions (''The Condition of Geometric Modular Action'') was
proposed in \cite{B/D/F/S}. One obtains directly transformation groups on
the indexed sets of the net. All these true QFT properties remain invisible
in any quantization approach. Combining modular theory with scattering
theory, the actual $J$ together with the incoming $J^{in}$ can be used to
obtain a new framework for nonperturbative interactions \cite{Schr1}. This
last topic will be treated in a separate paper by the present two authors.

Besides the old result of Bisognano/Wichmann there are nowadays other
interesting 'geometrical' actions of modular theory in chiral theories, see 
\cite{Bo/Yng} . We present some new examples in chiral and higher
dimensional models and give some outlook to future investigations of
geometrical actions.

\section{The Bisognano-Wichmann Property}

We will start with the famous result of Bisognano/Wichmann \cite{Bi/Wi}. We
will present a purely algebraic proof of that property.\footnote{%
After finishing this part we were informed by H.-J. Borchers of a recent
result by himself, \cite{Bo3}. Below we compare both approaches.}

Let us first introduce some notations. We assume given a Poincar\'{e}
covariant local net of observables which fulfills essential duality ( wedge
duality). Let $U$ denote the representation of the Poincar\'{e}' group on
the vacuum Hilbert space $\mathcal{H},$ $\Omega $ the vacuum vector. Let $%
l_{1}$ and $l_{2}$ be linear independent lightlike vectors and $%
W[l_{1}l_{2}] $ the wedge 
\begin{eqnarray}
W[l_{1},l_{2}]=\{\overrightarrow{x}\in R^{1,3}\mid &&\overrightarrow{x}%
=\alpha l_{1}+\beta l_{2}+l^{\perp },with \\
&&\alpha >0,\,\,\beta <0,\,(l^{\perp },l_{i})=(l^{\perp },l_{j})=0\} 
\nonumber
\end{eqnarray}
$\Lambda _{12}(t)$ the Lorentz boost to that wedge, i.e. the boosts which
map the wedge $W[l_{1},l_{2}]$ onto itself.

Let $\mathcal{A}(W[l_{1,}l_{2}])$ be the observable algebra associated to
the wedge. The Reeh-Schlieder property of the vacuum states that $\Omega $
is cyclic and separating for this algebra. We denote the associated modular
objects by $\Delta _{12},J_{12.}$

Before continuing with the results let us remind the reader on an
observation by Borchers \cite{Bo4}.

Assume \emph{wedge duality }for the underlying net. Then we have 
\begin{equation}
\Delta _{12}^{it}=F_{12}(t)U(\Lambda _{12}(-2\pi t))
\end{equation}
with some unitary group $F_{12}(t).$ Moreover modular theory gives 
\begin{equation}
\lbrack F_{12}(t),J_{12}]=[F_{12}(t),\Delta _{12}^{is}]=[F_{12}(t),U(\Lambda
_{12}(-2\pi s))]=0
\end{equation}
( $F_{12}$ leaves invariant $\mathcal{A}(W[l_{1,}l_{2}])$ and $\Omega .$)
Due to Borchers' result \cite{Bo1, Bo5} we immediately conclude 
\begin{equation}
\lbrack F_{12}(t),U(translations)]=0,
\end{equation}
see \cite{Bo4}. Moreover there is a strongly dense set of elements $A\in 
\mathcal{A}(W[l_{1},l_{2}])$ s.t. $U(\Lambda _{1,2}(2\pi t))A\Omega
,\,\,A\in \mathcal{A}(W[l_{1},l_{2}])$ can be analytically be continued to $%
t\in S(\frac{-1}{2},0)=\{z\in \mathbf{C\;}/\;-\frac{1}{2}<\func{Im}z<0\}..$

So far Borchers' results. Let us come to an observation which one of the
authors made in discussion with D. Guido:

\begin{theorem}
(Guido/Wiesbrock)

Additionally to the above we make the \emph{assumption } 
\begin{equation}
(A\Omega \longrightarrow U(\Lambda _{1,2}(-\frac{i}{2}))A^{*}\Omega
,\,\,\,A\in \mathcal{A}(W[l_{1},l_{2}]))
\end{equation}

$\,\,\,\,$ is uniformly bounded.

Then the Bisognano-Wichmann Property holds for the net.
\end{theorem}

%
%
%
%
%
%
%
%
%
%
\begin{proof}
Let $F_{12}(t)=e^{itf_{12}}$ be the unitary group above. From the assumption
it follows

\[
f_{12}\,\,\,\,\,\,\,\,\,\,\text{is semibounded.} 
\]

Now $F_{12}$ leaves invariant $\mathcal{A}(W[l_{1,}l_{2}])$ which implies by
a result of Borchers, \cite{Pe}, innerness of $F_{12,}$ i.e.

$\exists \,\,\,\,e^{itf}\in \mathcal{A}(W[l_{1,}l_{2}])$ with

\[
e^{itf}e^{-itf_{12}}\in \mathcal{A}(W[l_{1,}l_{2}])^{\prime }. 
\]

$J_{12}$-invariance of $F_{12}$ implies Ad $e^{itf_{12}}=$Ad $\
J_{12}e^{itf_{12}}J_{12}$ and therefore

\[
\ J_{12}e^{itf}e^{-itf_{12}}J_{12}=e^{i\lambda }e^{itf},\,\,\,\lambda \in 
\mathbf{R} 
\]

because $\mathcal{A}(W[l_{1,}l_{2}])$ is a factor. Putting together we
conclude

\[
e^{itf_{12}}=J_{12}e^{itf}J_{12}e^{itf}. 
\]

$e^{itf_{12}}$ is $\Delta _{12}$ invariant and therefore $e^{itf}$ too.
Similarly $e^{itf}$ is $U(\Lambda _{12})$ invariant and translation
invariant along the baseline of the wedge.

But due to the cluster property the later implies $e^{itf}=const.$
\end{proof}
%

We now make a comparison with the work of Borchers \cite{Bo3}.

As an algebraic criterion for the Bisognano-Wichmann Property, Borchers
proposed a so called reality condition as follows:

Let $A\in W[l_{1},l_{2}],$ such that $U(\Lambda _{1,2}(2\pi t))A^{*}\Omega
\; $can analytically be continued to $\func{Im}t=-\frac{1}{2}.$ (One can
find a *-strongly dense subset of such elements, see \cite{Bo4}) Then there
exists $\widehat{A}$ and $\widetilde{A}$ affiliated with $\mathcal{A}%
(W[l_{1},l_{2}])^{\prime }=\mathcal{A}(W[l_{2},l_{1}])$ with are related by: 
\begin{equation}
\medskip \medskip \widehat{A}\Omega =U(\Lambda _{1,2}(-i\pi ))A\Omega ,\;\;%
\widetilde{A}\Omega =U(\Lambda _{1,2}(-i\pi ))A^{*}\Omega .
\end{equation}
The reality condition states now:

\textit{For such }$A$\textit{\ one has \medskip }$\widehat{A}^{*}=\widetilde{%
A},$\textit{\ i.e. }$U(\Lambda _{1,2}(-i\pi ))A^{*}\Omega =\medskip \medskip 
\widehat{A}^{*}\Omega .$

Rewritten in terms of the modular data this means: 
\begin{equation}
U(\Lambda _{1,2}(-i\pi ))S\;A\Omega =S^{*}U(\Lambda _{1,2}(-i\pi ))\;A\Omega
\end{equation}
for a *-strongly dense subset of \medskip $\mathcal{A}(W[l_{1},l_{2}]),$
where $S$ denotes the Tomita conjugation to $(\mathcal{A}(W[l_{1},l_{2}]),%
\Omega ).\,$

Now due to commutativity 
\begin{equation}
T:=U(\Lambda _{1,2}(-i\pi ))\Delta ^{-\frac{1}{2}}=\Delta ^{-\frac{1}{2}%
}U(\Lambda _{1,2}(-i\pi ))
\end{equation}
is a s. a. operator and the reality condition can be rephrased as 
\begin{equation}
JTJ=T.
\end{equation}
But modular theory tells us that 
\begin{equation}
JU(\Lambda _{1,2}(-i\pi ))J=U(\Lambda _{1,2}(i\pi )),\,\,\,\,J\Delta ^{-%
\frac{1}{2}}J=\Delta ^{\frac{1}{2}}
\end{equation}
which implies $JTJ=T^{-1}$and therefore 
\begin{equation}
TT^{*}=T^{2}=TJTJ=TT^{-1}=1.
\end{equation}
Therefore the reality condition gives boundedness of $T$ by 1 which implies
the assumption of our theorem.

We don't see a natural physical motivation for our boundedness assumption.
But the results suggest that one should look at a purely algebraic criterion
which guarantees the Bisognano-Wichmann property. This would not rely on
some generating fields which in the algebraic approach to quantum field
theory play the role of a special choice of coordinating the physical system.

\section{On the modular theory of disjoint intervals}

In this section we present a modular interpretation of conformal
diffeomorphisms.

\subsection{The Virasoro-algebra{}}

We consider the $U(1)-$ current algebra on the circle. Then we have an
action of the Virasoro-algebra on the physical vacuum Hilbert space of our
model. The vacuum is defined as the unique invariant vector 
\begin{equation}
L_{1}\Omega =L_{-1}\Omega =L_{0}\Omega =0.
\end{equation}
The usual commutation relations are: 
\begin{equation}
\lbrack L_{n},L_{m}]=(n+m)L_{n+m}+\frac{1}{12}(m^{2}-1)m\delta _{m,n}
\end{equation}
so that: 
\begin{equation}
\lbrack L_{\pm 2},L_{0}]=\pm 2L_{\pm
2},\,\,\,\,\,\,\,\,\,[L_{2},l_{-2}]=4L_{0}+\frac{1}{2}
\end{equation}
holds. Then a simple computation shows 
\begin{eqnarray}
L_{0} &\mapsto &\frac{1}{2}L_{0}+\frac{1}{16} \\
L_{1} &\mapsto &\frac{1}{2\sqrt{2}}L_{2}  \nonumber \\
L_{-1} &\mapsto &\frac{1}{2\sqrt{2}}L_{-2}  \nonumber
\end{eqnarray}
gives an isomorphism of $sl(2\mathbf{R})-$ Lie algebras. This second
representation belongs to a $3-$dim. subgroup in Diff $S^{1}$ given as
follows. 
\begin{equation}
z\mapsto \sqrt{\frac{a+bz^{2}}{c+dz^{2}}\,}\,\,\,\,\,\,\,\,\,\,\,\left( 
\begin{array}{ll}
a & b \\ 
c & d
\end{array}
\right) \in SU(1,1)
\end{equation}
One notices the following (after discussion with M. Schmidt):

\begin{itemize}
\item  $SU(1,1)\,\,\,\,$maps the circle into itself

\item  the poles of $\frac{a+bz^{2}}{c+dz^{2}}$ lie outside the circle, the
zeros inside

\item  Choose the cuts by adjoining both zeroes and both poles. In that way
we get well defined maps.

\item  We get a representation of a twofold covering of the M\"{o}biusgroup.
\end{itemize}

Geometrically the associated dilations leave invariant disjoint intervals of
the following type: Given an interval, look at the square root of it
elements. We get a disjoint union of intervals as the inverse image. Both
parts separately are left invariant under that group. Therefore, if the
modular groups act like dilations as above, we do \textbf{not }have
Reeh-Schlieder Property for an arbitrary local algebra because of Takesaki's
result, see \cite{Stra}. I. e. this would either imply equality of the
algebra of a small interval and the algebra associated to appropriate
disjoint intervals or that the algebra is not cyclic.

But notice that due to the geometry it never happens that one part of such a
region covers both parts of another type of such disjoint intervals..
Especially we do not get into conflict with additivity of the net and
locality! This is the reason that such a modular theory is conceptually
allowed. Next we show that this also happens.

\subsection{Some Formulas}

\medskip Let us collect some well known formulas which help to understand
the following computations. In the $Sl(2\mathbf{R})$ group we have the
generator $D$ for the dilations, $P$ and $K$ for the translations resp.
conformal translations. They are related to the Virasoro group by 
\begin{eqnarray}
L_{0} &=&\frac{1}{2}(P+K)\,\,\,\,\,\,\,\,\,\,\,\,\,\,\,\,\,\,\,\,\,\,\,\,\,%
\,\,\,\,\,\,\,\,\,\,\,\,\,\,\,\,\,\,\,\,\,\,\,\,\,\,\,\,\,\,\,\,\,\,\,\,\,\,%
\,\,\,\,\,\,D=\frac{1}{2}(L_{1}-L_{-1})\,\,\,\,\,\,\,\,\,\,\,\,\,\,\,\,\,\,%
\,\,\,\,\,\,\,\,\,\,\,\,\,\,\,\,\,\,\,\,\,\,\,\,\, \\
L_{1} &=&D+\frac{1}{2}(P-K)\,\,\,\,\,\,\,\,\,\,\,\,\,\,\,\,\,\,\,\,\,\,\,\,%
\,\,\,\,\,\,\,\,\,\,\,\,\,\,\,\,\,\,\,\,\,\,\,\,\,\,\,\,\,P\,=\frac{1}{2}%
(L_{1}+L_{-1})\,\,+L_{0}  \nonumber \\
L_{-1} &=&D-\frac{1}{2}(P-K)\,\,\,\,\,\,\,\,\,\,\,\,\,\,\,\,\,\,\,\,\,\,\,\,%
\,\,\,\,\,\,\,\,\,\,\,\,\,\,\,\,\,\,\,\,\,\,\,\,\,\,\,\,\,\,\,K=\frac{1}{2}%
(L_{-1}+L_{1})\,\,-L_{0}.  \nonumber
\end{eqnarray}
It turns out to be helpful to switch from the compact (circle) picture to
the noncompact on the reals.

compact picture$\,\,\,\,\,\,\,\,\,\,\,\,\,\,\,\,\,\,\,\rightarrow
\,\,\,\,\,\,$non compact picture: $\,\,\,\,\,\,\,\,\,\,\,\,\,\,\,z\mapsto -i%
\frac{z-1}{z+1}$

non compact picture$\,\,\,\,\,\,\,\,\,\,\,\,\,\,\,\rightarrow \,\,\,\,\,$%
compact picture\thinspace :\thinspace \thinspace \thinspace \thinspace
\thinspace \thinspace \thinspace \thinspace \thinspace \thinspace \thinspace
\thinspace \thinspace \thinspace \thinspace \thinspace \thinspace \thinspace
\thinspace \thinspace \thinspace \thinspace \thinspace $x\mapsto \,\,\frac{%
ix+1}{-ix+1}.$

Under these transformation the relevant forms and the $L_{n}$ are mapped to
the following. 
\begin{eqnarray}
dx &=&-i\frac{2}{\left( z+i\right) ^{2}}dz,\,\,\,dz=2i\frac{1}{\left(
-ix+1\right) ^{2}}dx \\
L_{n} &=&z^{n+1}\frac{d}{dz},\,\,\,\,L_{n}=\frac{\left( -ix+1\right) ^{2}}{2i%
}\left( \frac{ix+1}{-ix+1}\right) ^{n+1}\frac{d}{dx}  \nonumber
\end{eqnarray}
A formal computation now shows:

$z^{2}\circ L_{n}\circ \sqrt{z}=\frac{1}{2}L_{2n},$ i.e.\thinspace
\thinspace :\thinspace \thinspace \thinspace \thinspace \thinspace
\thinspace \thinspace 
\begin{eqnarray}
L_{\pm 1} &\mapsto &\,\,\,\,\,\frac{1}{2}L_{\pm 2} \\
L_{0} &\mapsto &\,\,\,\,\,\frac{1}{2}L_{0}  \nonumber
\end{eqnarray}
in the non compact picture:

\begin{eqnarray}
z\,\,\, &\mapsto &\,\,z^{2}\,\,\,\,\simeq \,\,x\mapsto \,\,\frac{2x}{1-x^{2}}
\\
z\,\,\, &\mapsto &\,\,\sqrt{z}\,\,\simeq \,\,x\mapsto \,\,-\frac{1}{x}%
+sign(x)\sqrt{1+(\frac{1}{x})^{2}}\,\,\,  \nonumber
\end{eqnarray}
$\,\,\,\,\,\,\,\,\,\,\,\,\,\,\,\,\,\,\,\,\,\,\,\,\,\,\,\,\,\,\,\,\,\,\,\,\,%
\,\,\,\,$

In the non compact and compact picture it is easily seen that
diffeomorphisms act symplectically: 
\begin{equation}
\omega (g,f)=\frac{1}{2}\int g(x)df(x),\,\,\,\,\,\,\,\omega (g,f)=\frac{1}{2}%
\int g(z)df(z)
\end{equation}

\subsection{\protect\medskip Transformation and Invariant Scalar Product}

Let $\left( 
\begin{array}{ll}
a & b \\ 
c & d
\end{array}
\right) \in Sl(2\mathbf{R}),$ then the underlying symmetry action is: 
\begin{equation}
x\mapsto \;-\frac{2cx+d(1-x^{2})}{2ax+b(1-x^{2})}+sign(x)\sqrt{1+\left( -%
\frac{2cx+d(1-x^{2})}{2ax+b(1-x^{2})}\right) ^{2}}.
\end{equation}

Moreover

\begin{equation}
<f,g>=\int \frac{f(x)g(y)}{\left( (x-y)(1+xy)+i0\right) ^{2}}%
(1+x^{2})(1+y^{2})dxdy
\end{equation}

gives an invariant scalar product.

\textbf{proof:}

Denote $\Gamma (x)=-\frac{1}{x}+sign(x)\sqrt{1+(\frac{1}{x})^{2}}\,\,\,$then
we compute 
\begin{eqnarray}
&&\int \frac{f\circ \Gamma (\frac{a(\frac{2x}{1-x^{2}})+b}{c(\frac{2x}{%
1-x^{2}})+d})g\circ \Gamma (\frac{a(\frac{2y}{1-y^{2}})+b}{c(\frac{2y}{%
1-y^{2}})+d})}{((x-y)(1+xy)+i0)^{2}}(1+x^{2})(1+y^{2})dxdy \\
&=&\int \frac{f\circ \Gamma (\frac{a(\frac{2x}{1-x^{2}})+b}{c(\frac{2x}{%
1-x^{2}})+d})g\circ \Gamma (\frac{a(\frac{2y}{1-y^{2}})+b}{c(\frac{2y}{%
1-y^{2}})+d})}{(\frac{2x}{1-x^{2}}-\frac{2y}{1-y^{2}}+i0)^{2}}\frac{%
2(1+x^{2})2(1+y^{2})}{(1-x^{2})^{2}(1-y^{2})^{2}}dxdy  \nonumber
\end{eqnarray}
One notices $d(\frac{2x}{1-x^{2}})=\frac{2(1+x^{2})}{(1-x^{2})^{2}}$ and
thereby: 
\[
=\int \frac{f\circ \Gamma (\frac{a(\frac{2x}{1-x^{2}})+b}{c(\frac{2x}{1-x^{2}%
})+d})g\circ \Gamma (\frac{a(\frac{2y}{1-y^{2}})+b}{c(\frac{2y}{1-y^{2}})+d})%
}{(\frac{2x}{1-x^{2}}-\frac{2y}{1-y^{2}}+i0)^{2}}d(\frac{2x}{1-x^{2}})d(%
\frac{2y}{1-y^{2}})
\]
\begin{eqnarray}
&=&\int \frac{f\circ \Gamma (\frac{2x}{1-x^{2}})g\circ \Gamma (\frac{2y}{%
1-y^{2}})}{(\frac{2x}{1-x^{2}}-\frac{2y}{1-y^{2}}+i0)^{2}}d(\frac{2x}{1-x^{2}%
})d(\frac{2y}{1-y^{2}}) \\
&=&\int \frac{f(x)g(y)}{((x-y)(1+xy)+i0)^{2}}(1+x^{2})(1+y^{2})dxdy 
\nonumber
\end{eqnarray}
where the last equality follows from a substitution with $(x\mapsto \frac{2x%
}{1-x^{2}})$ , which establishes invariance. qed.

The form is positive definite: Write $\,f=f\circ \Gamma \circ (\frac{2x}{%
1-x^{2}}).$

Define 
\begin{equation}
I(f)(x)=\int f(y)K_{I}(x,y)dy
\end{equation}
with 
\begin{equation}
k_{I}(x,y)=\frac{-1}{(x-y)}-\frac{1}{y}\frac{1}{(1+xy)}=\frac{-x(1+y^{2})}{%
y(x-y)(1+xy)}.
\end{equation}
Then one computes: 
\begin{equation}
\frac{d}{dx}k_{I}(x,y)=\frac{(1+x^{2})(1+y^{2})}{[(x-y)(1+xy)]^{2}}=\frac{-1%
}{(x-y)^{2}}-\frac{1}{(1+xy)^{2}}.
\end{equation}
and we get: 
\begin{equation}
\omega (If,g)=<f,g>=-\omega (f,Ig).
\end{equation}
Next we show that this scalar product gives us a Fock state on the Weyl
algebra:

Denote 
\begin{equation}
\Gamma (f)(x):=f(\frac{x}{2}+sign(x)\sqrt{1+\left( \frac{x}{2}\right) ^{2}})
\end{equation}
and 
\begin{equation}
I_{0}(f)(x)=\int \frac{-1}{(x-y+i0)}f(y)dy
\end{equation}
\medskip which gives the usual vacuum Fock state of the theory \cite{Yng}. 
\[
\left\langle f,g\right\rangle _{0}=\omega (I_{0}f,g)=\int \frac{f(x)g(y)}{%
(x-y+i0)^{2}}dxdy 
\]

Then we compute: 
\begin{eqnarray}
\Gamma ^{-1}\circ I_{0}\circ \Gamma (f)(x) &=&\int \frac{-1}{(-\frac{1}{x}%
+x-y+i0)}f(\frac{y}{2}+sign(x)\sqrt{1+\left( \frac{y}{2}\right) ^{2}})dy \\
&=&\int \frac{-1}{(-\frac{1}{x}+x-(-\frac{1}{z}+z)+i0)}f(z)\frac{z^{2}+1}{%
z^{2}}dz  \nonumber \\
&=&\int \frac{-x(z^{2}+1)}{z^{2}(-1+x^{2}+\frac{x}{z}-zx+i0)}f(z)dz 
\nonumber \\
&=&\int \frac{-x(z^{2}+1)}{z(-z+x^{2}z+x-z^{2}x+i0)}f(z)dz  \nonumber \\
&=&\int \frac{-x(z^{2}+1)}{z(-z+x+i0)(1+xz)}f(z)dz=I(f)(x)  \nonumber
\end{eqnarray}
( One has to take care of the $+i0$-term?!), so we get: 
\begin{equation}
\Gamma ^{-1}\circ I_{0}\circ \Gamma =I.
\end{equation}
As is well known $I_{0}^{2}=-1,$ from which we immediately conclude $%
I^{2}=-1,$ i.e. the invariant scalar product produces a Fock state on the
Weyl algebra.

\subsection{The KMS-Condition}

Let us show the KMS-Condition. For this we first make the following simple
observation: 
\begin{eqnarray}
f &\in &C_{0}^{\infty }([0,1])\Rightarrow \Gamma (f)\in C_{0}^{\infty
}([0,\infty [)  \nonumber \\
f &\in &C_{0}^{\infty }(]-\infty ,-1])\Rightarrow \Gamma (f)\in
C_{0}^{\infty }([0,\infty [).
\end{eqnarray}
Denote $<,>_{0}$ the scalar product of the vacuum state, $<,>_{1}$ the one
above. Then we show: 
\[
<\Gamma (f),\Gamma (g)>_{0}=<f,g>_{1}: 
\]
\begin{eqnarray}
&&\int dxdy\frac{f(-\frac{1}{x}+sign(x)\sqrt{1+(\frac{1}{x})^{2}})g(-\frac{1%
}{y}+sign(x)\sqrt{1+(\frac{1}{y})^{2}})}{(x-y+i0)^{2}}  \nonumber \\
&=&\int dzdw\frac{f(z)g(w)}{(\frac{2z}{1-z^{2}}-\frac{2w}{1-w^{2}}+i0)^{2}}%
\frac{1+z^{2}}{(1-z^{2})^{2}}\frac{1+w^{2}}{(1-w^{2})^{2}}  \nonumber \\
&=&\frac{1}{2}\int dzdw\frac{f(z)g(w)}{(z-w+i0)^{2}}\frac{(1+z^{2})(1+w^{2})%
}{(1+zw)^{2}}
\end{eqnarray}
Let us also introduce the notion 
\begin{equation}
V(\lambda )(f)(x)=f(\lambda x),\,\,\,\,\,\,\,\,\,\,\,\,\,U(\lambda )(f)(x)=f(%
\frac{1-x^{2}}{2\lambda ^{2}}+sign(x)\sqrt{1+\left( \frac{1-x^{2}}{2\lambda
^{2}}\right) ^{2}})
\end{equation}
then, an easy computation shows 
\begin{equation}
\Gamma ^{-1}\circ V(\lambda )\circ \Gamma =U(\lambda ).
\end{equation}
To prove the KMS-property for the disjoint interval $]-\infty ,-1]\cup [0,1]$
w.r.t. $U,$ we use the strategy as in \cite{Yng} .

Let $f,g\in C_{0}^{\infty }([0,1]).$ Then 
\begin{equation}
<U(\lambda )f,g>_{1}=<V(\lambda )\circ \Gamma (f),\Gamma (g)>_{0}
\end{equation}
Due to (3.1), $\Gamma (f),\Gamma (g)\in C_{0}^{\infty }([0,\infty [)$. J.
Yngavson has shown in \cite{Yng} that for such functions the expectation
value with $V(\lambda )$ can analytically be continued to $\func{Im}\lambda
=2\pi $ with specific boundary values. This gives in our case: 
\begin{eqnarray}
\stackunder{\theta \uparrow 2\pi }{\lim } &<&U(\lambda )f,g>_{1}=\stackunder{%
\theta \uparrow 2\pi }{\lim }<V(\lambda )\circ \Gamma (f),\Gamma
(g)>=\left\langle \Gamma (g),\Gamma (f)\right\rangle = \\
&=&\left\langle g,f\right\rangle _{1}  \nonumber
\end{eqnarray}
This shows that $U(\lambda )$ fulfils the KMS-condition for the Weyl-algebra
to $[0,1].$

By the invariance and relation (3.2) we immediately get the same result for $%
f,g\in C_{0}^{\infty }(]-\infty ,-1]),$ resp. $f\in C_{0}^{\infty }(]-\infty
,-1]),\,g\in C_{0}^{\infty }([0,1]).$and thereby get that $U\,\,$gives the
modular group to the disjoint intervals. The methods of \cite{Yng} then also
apply to give$\;$%
\begin{equation}
J=\Gamma ^{-1}\circ J_{0}\circ \Gamma \Rightarrow J(f)(x)=f^{*}(\frac{1}{x})
\end{equation}
\medskip The same methods should also work for the $L_{3,}L_{-3}$ case,
where one has to use the third root. In this way we get an action of these
other diffeomorphism groups as automorphisms on the local net. As is well
known they can be implemented on the vacuum theory by Bogoliubov
automorphisms.

We expect that these results hold in all loop group models as they were
constructed for example in \cite{Was}.

The final upshot of this section is that several diffeomorphism actions can
occur as the action of some modular groups of certain observable algebras
w.r.t. to a proper state.

\section{Hidden Symmetry}

As was shown in \cite{Bo/Yng} it might happen that the modular group acts
only geometrically on a certain subregion. We show how this comes up very
naturally in higher dimensional situations. We start with some purely group
theoretical observations on an interesting subgroup of the Poincar\'{e}
group \cite{Schr3}. The standard wedge situation suggests to decompose the
Poincar\'{e} group generators into longitudinal, transversal and mixed
generators 
\begin{equation}
\,P_{\pm }=\frac{1}{\sqrt{2}}(P_{t}\pm
P_{z}),\,\,M_{tz};\,\,M_{xy},\,\,P_{i};\,\,G_{i}^{(\pm )}\equiv \frac{1}{%
\sqrt{2}}(M_{it}\pm M_{iz}),\,i=x,y
\end{equation}
The generators $G_{i}^{(\pm )}$ are precisely the ``translational'' pieces
of the euclidean stability groups $E^{(\pm )}(2)$ of the two light vectors $%
e^{(\pm )}=(1,0,0,\pm 1)$ which appeared for the first time in Wigner's
representation theory for zero mass particles. As one reads off from the
commutation relations, $P_{i},G_{i}^{(+)},P_{\pm }$ have the interpretation
of a central extension of a transversal ``Galilei group'' with the two
``translations'' $G_{i}^{(+)}$ representing the Galilei generators, $P_{+}$
the central ``mass'' and $P_{-}$ the ``nonrelativistic Hamiltonian''. The
longitudinal boost $M_{tz}$ scales the Galilei generators $G_{i}^{(+)}$ and
the ``mass'' $P_{+}.$ Geometrically the $G_{i}^{(+)}$ change the standard
wedge (it tilts the longitudinal plane) and the corresponding finite
transformations generate a family of wedges whose envelope is the half-space 
$x_{-}\geq 0.$ The Galilei group together with the boost $M_{tz}$ generate
an 8-parametric subgroup $G^{(+)}(8)$ inside the 10-parametric Poincar\'{e}
group: 
\begin{equation}
\,G^{(+)}(8):\,\,P_{\pm },\,\,M_{tz};\,\,M_{xy},\,\,P_{i};\,\,G_{i}^{(+)}
\label{gen}
\end{equation}
The modular reflection $J$ transforms this group into an isomorphic $%
G^{(-)}(8).$ All observation have interesting generalizations to the
conformal group in massless theories in which case the associated natural
space-time region is the double cone. This subgroup is intimately related to
the notion of ``modular intersection''. Let $l_{1},l_{2}$ and $l_{3}$ be 3
linear independent light-like vectors and consider two wedges $%
W(l_{1},l_{2}),W(l_{1},l_{3})$ with $\Lambda _{12}$ and $\Lambda _{13}$ the
associated Lorentz boosts. As a result of this common $l_{1}$ the algebras $%
\mathcal{N}=\mathcal{A}(W(l_{1},l_{2})),\mathcal{M}=\mathcal{A}%
(W(l_{1},l_{3}))$ have a modular intersection with respect to the vector $%
\Omega .$ Especially ($\mathcal{N\cap M})\subset \mathcal{M},\Omega )$ is a
modular inclusion \cite{Wi1,Ar/Zs,Bo5}. Identifying $W(l_{1},l_{2})$ with
the above standard wedge, we notice that the longitudinal generators $P_{\pm
},\,\,M_{tz}$ are related to the inclusion of the standard wedge algebra
into the full algebra $B(H),$ whereas the Galilei generators $G_{i}^{(+)}$
are the ``translational'' part of the stability group of the common light
vector $l_{1}$ (i.e. of the Wigner light-like little group). These
generators $G_{i}^{(+)}$ should be considered as being associated with a
common light ray shared by two wedges whereas all the other generators in (%
\ref{gen}) are either longitudinal or transversal to one wedge (the standard
wedge).

To simplify the situation let us take d=1+2 with $G^{(+)}(4),$ in which case
there is only one Galilei generator $G.\,$ In addition to the ``visible''
geometric subgroup of the Poincar\'{e} group, the modular theory produces a
``hiddensymmetry transformation which belongs to a region which is a
intersection of two wedges.

Let us now consider the 3-dim. situation from the point of view of algebraic
QFT. We assume the Bisognano-Wichmann property for the net and also assume
that it is fulfills additivity. Denote $l_{1,}l_{2},l_{3}$ three linear
independent lightlike vectors and $W[l_{1,}l_{2}]$ , $W[l_{1,}l_{2}]$ two
wedges characterized by their two lightlike vectors. Denote $\Lambda _{12}$
resp. $\Lambda _{13}$ the related Lorentz boosts. Due to their common
light-ray $l_{1}$ the associated observable algebras have - modular
intersection w.r.t. $\Omega $, see \cite{Bo2,Wi4}. To simplify the notation
we use $\mathcal{N}$\emph{,}$\mathcal{M}$\emph{\ }$~$~for the algebras. Then 
$((\mathcal{N}\cap \mathcal{M)}\subset \mathcal{M},\Omega )\,$is a modular
inclusion, see \cite{Wi1, Ar/Zs,Bo2,Wi4} and 
\begin{equation}
U_{\mathcal{N\cap M},\mathcal{M}}(a):=\exp (\frac{ia}{2\pi }(\ln \Delta
_{N\cap M}-\ln \Delta _{M}))
\end{equation}
is a unitary group with positive generator. Moreover one has 
\begin{equation}
U_{\mathcal{N\frown M},\mathcal{M}}(1-e^{-2\pi t})=\Delta _{\mathcal{M}%
}^{it}\Delta _{\mathcal{N\frown M}}^{-it}
\end{equation}
\begin{equation}
U_{\mathcal{N\cap M},\mathcal{M}}(e^{-2\pi t}a)=\Delta _{\mathcal{M}}^{it}U_{%
\mathcal{N\cap M},\mathcal{M}}(a)\Delta _{\mathcal{M}}^{-it}
\end{equation}
\begin{equation}
AdU_{\mathcal{N\cap M},\mathcal{M}}(-1)(\mathcal{M}\emph{)=}\mathcal{N}\cap 
\mathcal{M}
\end{equation}
and 
\begin{equation}
J_{\mathcal{M}}U_{\mathcal{N\frown M},\mathcal{M}}(a)J_{\mathcal{M}}=U_{%
\mathcal{N\frown M},\mathcal{M}}(-a).
\end{equation}
Similar results hold for $\mathcal{N}$\emph{\ }replacing $\mathcal{M}$ , see 
\cite{Bo2,Wi3}\emph{. }Due to the intersection property we finally have the
relation
\[
\lbrack U_{\mathcal{N\frown M},\mathcal{M}}(a),U_{\mathcal{N\frown M},%
\mathcal{N}}(b)]=0
\]
which enables one to define the unitary group
\[
U_{\mathcal{N\frown M}}(a)=U_{\mathcal{N\frown M},\mathcal{M}}(-a)U_{%
\mathcal{N\frown M},\mathcal{N}}(a).
\]
This later group can be rewritten as
\[
U_{\mathcal{N\frown M}}(1-e^{-2\pi t})=\Delta _{\mathcal{M}}^{it}\Delta _{%
\mathcal{N}}^{-it}
\]
and thereby recognized to be in our physical application the 1-parameter
Galilean subgroup $G$ (\ref{gen}) in the above remarks.

Now we notice that for $a<0$%
\begin{eqnarray}
AdU_{\mathcal{N\frown M},\mathcal{M}}(a)(\mathcal{M}\emph{)} &=&Ad\Delta _{%
\mathcal{M}}^{-i(\frac{1}{2\pi }\ln -a)}U_{\mathcal{N\frown M},\mathcal{M}%
}(-1)(\mathcal{M}) \\
&=&Ad\Delta _{\mathcal{M}}^{-i(\frac{1}{2\pi }\ln -a)}(\mathcal{N}\cap 
\mathcal{M})  \nonumber
\end{eqnarray}
Because $\Delta _{\mathcal{M}}^{it}$ acts geometrically as Lorentz boosts,
we fully know the geometrical action of $U_{\mathcal{N\frown M},\mathcal{M}%
}(a)$ on $\mathcal{M}$\emph{\ }for\emph{\ }$a<0.$ For $~a>0$ we notice 
\begin{eqnarray}
AdU_{\mathcal{N\frown M},\mathcal{M}}(1)(\mathcal{M}\emph{)} &=&AdU_{%
\mathcal{N\frown M},\mathcal{M}}(2)(\mathcal{M}\cap \mathcal{N}\emph{)}=AdJ_{%
\mathcal{M}}J_{\mathcal{N\frown M}}(\mathcal{M}\cap \mathcal{N}\emph{)} \\
&=&AdJ_{\mathcal{M}}(\mathcal{M}^{\prime }\cup \mathcal{N}^{\prime }) 
\nonumber
\end{eqnarray}
and again, due to the geometrical action of $J_{M}$ we have a geometrical
action on $\mathcal{M}$ for $a>0.$%
\begin{equation}
AdU_{\mathcal{N\cap M},\mathcal{M}}(a)(\mathcal{M}\emph{)=}Ad\Delta _{%
\mathcal{M}}^{-i(\frac{1}{2\pi }\ln a)}J_{\mathcal{M}}(\mathcal{M}^{\prime
}\cup \mathcal{N}^{\prime })
\end{equation}
From these observations and with $U_{\mathcal{N\cap M},\mathcal{M}%
}(1-e^{-2\pi t})=\Delta _{\mathcal{M}}^{it}\Delta _{\mathcal{M\cap N}}^{-it}$
we get for $t<0:$%
\begin{equation}
Ad\Delta _{\mathcal{N\cap M}}^{it}(\mathcal{M})=Ad\Delta _{\mathcal{M}}^{(-%
\frac{i}{2\pi }\ln (e^{-2\pi t}-1))}J_{\mathcal{M}}(\mathcal{M}^{\prime
}\cup \mathcal{N}^{\prime })
\end{equation}
and in case of $t>0:$%
\begin{equation}
Ad\Delta _{\mathcal{N\cap M}}^{it}(\mathcal{M})=Ad\Delta _{\mathcal{M}}^{(-%
\frac{i}{2\pi }\ln (1-e^{-2\pi t}))}(\mathcal{N}\cap \mathcal{M}).
\end{equation}
Similar results hold for $\mathcal{N}$\emph{\ }~replacing $\mathcal{M}$\emph{%
\ . }With the same methods we get:\smallskip 
\begin{eqnarray}
Ad\Delta _{\mathcal{N\cap M}}^{it}\Delta _{\mathcal{N}}^{is}(\mathcal{M})
&=&Ad\Delta _{\mathcal{N\cap M}}^{it}\Delta _{\mathcal{N}}^{is}\Delta _{%
\mathcal{M}}^{-is}(\mathcal{M}) \\
&=&Ad\Delta _{\mathcal{N\cap M}}^{it}U_{\mathcal{M\cap N}}(e^{-2\pi s}-1)(%
\mathcal{M})  \nonumber
\end{eqnarray}
where $U_{\mathcal{N\cap M}}$ is the 1-parameter Lorentz subgroup (the
Galilei subgroup $G$ in (\ref{gen}) associated with the modular
intersection, see \cite{Bo2,Wi4}. This gives: 
\begin{eqnarray}
Ad\Delta _{\mathcal{N\cap M}}^{it}\Delta _{N}^{is}(\mathcal{M}) &=&AdU_{%
\mathcal{M\cap N}}(e^{-2\pi t}(e^{-2\pi s}-1))\Delta _{\mathcal{N\cap M}%
}^{it}(\mathcal{M}) \\
&=&AdU_{\mathcal{M\cap N}}(e^{-2\pi t}(e^{-2\pi s}-1))\Delta _{\mathcal{M}%
}^{-\frac{1}{2\pi }\ln (1-e^{-2\pi t})}(\mathcal{M\cap N}),  \nonumber
\end{eqnarray}
if $t>0$ and similar for $t<0$.Therefore we get a geometrical action of $%
\Delta _{\mathcal{N\cap M}}^{it}$ on $Ad\Delta _{\mathcal{N}}^{is}(\mathcal{M%
}).$

A look at the proof shows that the essential ingredients are the special
commutation relations. Due to
\[
\Delta _{\mathcal{M\cap N}}^{it}=\Delta _{\mathcal{M}}^{it}U_{\mathcal{N\cap
M},\mathcal{M}}(1-e^{-2\pi t})=\Delta _{\mathcal{M}}^{it}J_{\mathcal{M}}U_{%
\mathcal{N\cap M},\mathcal{M}}(e^{-2\pi t}-1)J_{\mathcal{M}}
\]
and the well established geometrical action of $\Delta _{\mathcal{M}}^{it}$
and $J_{\mathcal{M}}$ it is enough to consider the action of $U_{\mathcal{%
N\cap M},\mathcal{M}}$ or similarly $U_{\mathcal{N\cap M},\mathcal{N}}.$ For
these groups we easily get
\[
AdU_{\mathcal{N\cap M},\mathcal{M}}(a)\Delta _{\mathcal{N}}^{is}\Delta _{%
\mathcal{M}}^{-it}(\mathcal{N})=Ad\Delta _{\mathcal{N}}^{is}\Delta _{%
\mathcal{M}}^{it}U_{\mathcal{N\cap M},\mathcal{M}}(e^{-2\pi (s+t)}a)(%
\mathcal{N)}
\]
and due to the above remarks the geometrical action of  $\Delta _{\mathcal{%
N\cap M}}^{it}$ on the algebras of the type $Ad\Delta _{\mathcal{N}%
}^{is}\Delta _{\mathcal{M}}^{-it}(\mathcal{M})$.

Now, the lightlike translations $U_{transl_{1}\,}(a)$ in $l_{1}$ direction
fulfill the positive spectrum condition and map $\mathcal{N}\cap \mathcal{M}$
into itself for $a>0.$ Therefore we have the Borchers commutator relations
with $\Delta _{\mathcal{M\cap N}}^{it}$ and get 
\begin{equation}
Ad\Delta _{\mathcal{N\frown M}}^{it}U_{transl_{1}}(a)(\mathcal{M}%
)=AdU_{transl_{1}}(e^{-2\pi t}a)\Delta _{\mathcal{N\frown M}}^{it}(\mathcal{M%
})
\end{equation}
The additivity of the net tells us that taking unions of the algebra
corresponds to the causal unions of localization regions. The assumed
duality allows us to pass to causal complements and thereby to intersections
of the underlying localization regions. Therefore the algebraic properties
above transfer to unions, causal complements and intersections of regions.
We finally get 

\begin{theorem}
Let $\mathcal{G}$ be the set of regions in $\mathbf{R}^{1,2}$ containing the
wedges $W[l_{1},l_{2}],W[l_{1},l_{3}]$ and which is closed under

a) Lorentz boosting with $\Lambda _{12}(t),\Lambda _{13}(s),$

b) intersection

c) (causal) union

d) translation in $l_{1}$ direction

e) causal complement

Then $\Delta _{W[l_{1},l_{2}]\cap W[l_{1},l_{3}]}^{it}$ maps sets in $%
\mathcal{G}$ onto sets in $\mathcal{G}$ in a well computable way and extends
the subgroup (\ref{gen}) by a ``hidden symmetry''.
\end{theorem}

Similarly we can look at a (1+3)-dim. quantum field theory. Then we get the
same results as above for the modular theory to the region $%
W[l_{1},l_{2}]\cap W[l_{1},l_{3}]\cap W[l_{1},l_{4}],$ where $l_{i}$ are 4
linear independent lightlike vectors in $\mathbf{R}^{1,3}.$ Moreover in this
case the set $\mathcal{G}$ contain $W[l_{1},l_{2}],W[l_{1},l_{3}]$ and $%
W[l_{1},l_{4}]$ and is closed under boosting with $\Lambda _{12}(t),\Lambda
_{13}(s),\Lambda _{14}(r).$

The arguments are based on the Borchers commutation relation and modular
theory and apply also if we replace modular intersection by modular
inclusion. One recovers in this way the results of Borchers and Yngvason, 
\cite{Bo/Yng}. ( Note that in thermal situations we have no simple
geometrical interpretation for the commutants as the algebra to causal
complements. Therefore in these cases we have to drop e) in the above
theorem.).

The final upshot of this section is to show that there might be sensible
meanings of geometrical actions of modular groups by restricting on certain
subsystems.

\section{4-dim. Theories from a Finite Set of Algebras}

Let us mention a recent result due to K\"{a}hler and one of the authors
(H.-W. W). It follows a line beginning with the work of Borchers, \cite{Bo1}
and one of the authors, \cite{Wi2}. Starting with a finite set of algebras
lying in a specified position w.r.t. their common modular theory one
constructs a net of local observables.

\begin{theorem}
Let $\mathcal{M}_{ij},\,\,\,0\leq i<j\leq 4,\,\,\mathcal{M}_{ij}^{\prime }=%
\mathcal{M}_{ji},\,$ be $\,$von-Neumann algebras acting on~$~\mathcal{H}$, $%
\Omega \,\,\,\in \mathcal{H}$ $~\,$with: 
\begin{equation}
a)\,\,\,\,\,\,(\mathcal{M}_{ij},\mathcal{M}_{ik},\Omega )\,\,\,\,\,\,\mathsf{%
\,}has\,\,\,modular\,\,intersection
\end{equation}
\emph{This part reflects the underlying wedge geometry.}
\end{theorem}

\[
b)\,\,\,\,\,\,symmetric\,\,in\,\,indices 
\]
\emph{This part means that there is no preferential ''wedge'' region. We can
reformulate this as\thinspace \thinspace follows:}

Let $\Gamma _{P}:=\Delta _{24}^{it_{0}}\Delta _{34}^{-it_{0}}\Delta
_{32}^{it_{0}}\Delta _{42}^{-it_{0}}$with: 
\begin{equation}
b1)\,\,\,\,\,\,\,\,\,\,\,\,\,\,\,\,\,Ad\,\,\,\Gamma _{P}(\mathcal{M}_{14})=%
\mathcal{M}_{12},\,\,\,\,\,\,\,\,\,Ad\,\,\,\Gamma _{P}(\mathcal{M}_{13})=%
\mathcal{M}_{14}
\end{equation}
and $\Gamma _{P^{\prime }}:=\Delta _{13}^{it_{0}}\Delta
_{43}^{-it_{0}}\Delta _{14}^{-it_{0}}\,\,\Delta _{13}^{-t_{0}}\,$with 
\begin{equation}
b2)\,\,\,\,\,\,\,\,\,\,\,\,\,Ad\,\,\,\Gamma _{P^{\prime }}(\mathcal{M}_{12})=%
\mathcal{M}_{32},\,\,\,\,\,\,\,\,\,Ad\,\,\,\Gamma _{P}(\mathcal{M}_{42})=%
\mathcal{M}_{12}
\end{equation}
\[
c)\,\,\,closed\,\,in\,\,some\,\,finite\,\,\dim .\,\,Lie\,\,group 
\]
\emph{Essentially we use that the generators of the modular groups can be
composed. We reformulate this as follows.}

Denote $P^{1,1}$ the group generated by \medskip $\{\Delta
_{14}^{it_{14}}\Delta _{24}^{it_{24}}\Delta _{34}^{it_{34}}\}$ , $SO(1,2)$
the one generated by$\,\,\,\{\Delta _{13}^{it_{13}}\Delta
_{12}^{it_{12}}\Delta _{23}^{it_{23}}\}\,\,$Then let 
\begin{equation}
c^{\prime })\,\,\,\,\,\,\,Ad\,\,J_{12}(P_{\varepsilon }^{1,1})\subset
P_{\delta }^{1,1}\cdot SO(1,2),\,\,\,P_{\varepsilon }^{1,1},\,\,\varepsilon
-neighborhood\,\,of\,\,1\in \text{ }P^{1,1}
\end{equation}
Then we conclude: 
\begin{equation}
\left\{ \Delta _{14}^{it_{14}},\Delta _{24}^{it_{24}},\Delta
_{34}^{it_{34}}\right\} ,\left\{ \Delta _{12}^{it_{12}},\Delta
_{13}^{it_{13}},\Delta _{23}^{it_{23}}\right\}
\,\,\,generate\,\,a\,\,reprs.\,\,of\,\,Sl(2\mathbf{C).}
\end{equation}

\begin{proof}
see \cite{Ka/Wi}
\end{proof}

In order to get a representation of the Poincar\'{e} group we use the
following. First we again implement rudimentarily the wedge geometry by:

Let $\mathcal{N}_{12}\,$be a von-Neumann Algebra $\,$with 
\begin{equation}
d)\,\,\,\,\,\,(\mathcal{N}_{12}\subset \mathcal{M}_{12},\Omega
)\,\,\,is\,\,\,hsm.
\end{equation}
The resulting translations should be reflected by the CPT-like conjugations: 
\begin{equation}
e1)\,\,\,\,\,\,\,\,\,\,\,\,\,Ad\,\,J_{\mathcal{M}_{1j}}(J_{\mathcal{N}%
_{12}}J_{\mathcal{M}_{12}})=J_{\mathcal{M}_{12}}J_{\mathcal{N}%
_{12}},\,\,\,\,\,\,\,\,\,\,\,\,\,\,\,\,~j=3,4
\end{equation}
\begin{equation}
e2)\,\,\,\,\,[\text{Ad}\,\,J_{\mathcal{M}_{jk}}(J_{\mathcal{N}_{12}}J_{%
\mathcal{M}_{12}}),J_{\mathcal{N}_{12}}J_{\mathcal{M}_{12}}]=0\,\,,\,\,\,\,%
\,\,\,\,\,\,\,\,\,\,\,\,\,\,\,\,\,\,j,k=2,3,4
\end{equation}
and the symmetry in the indices (no preferential wedge) leads to: 
\begin{equation}
f)\,\,\,\,\,Ad\,\,\,\Gamma _{P}(J_{\mathcal{N}_{12}}J_{\mathcal{M}_{12}})=J_{%
\mathcal{N}_{12}}J_{\mathcal{M}_{12}}
\end{equation}

\begin{remark}
Notice that Ad $\Delta _{12}^{it_{0}}\Delta _{13}^{-it_{0}}\Gamma _{P}\,(%
\mathcal{M}_{12})=\mathcal{M}_{12},$ so that f) implies
\end{remark}

Ad$\Delta _{12}^{it_{0}}\Delta _{13}^{-it_{0}}\Gamma _{P}(J_{\mathcal{N}%
_{12}})=J_{\mathcal{N}_{12}}.$ Modular theory then shows Ad $\Delta
_{12}^{it_{0}}\Delta _{13}^{-it_{0}}\Gamma _{P}\,(\mathcal{N}_{12})=\mathcal{%
N}_{12}.$

Then we get a representation of $\mathbf{R}^{1,3}$ , the translations.

The Lorentz group maps translations onto themselves which can be encoded in: 
\begin{equation}
g)\,\,\,Ad\,\,J_{jk}(translations)\subset translations
\end{equation}
Under these assumptions we get a representation of the full Poincar\'{e}
group with spectrum condition. Using this one easily constructs a net of
observables in $\mathbf{R}^{1,3}$ to it.

These results show that spacetime can be encoded in a finite set of
algebraic data. Moreover spacetime is recovered by looking at the
noncommutative structure measured w.r.t. the underlying physical state of
the system (modular theory). This would not work in a classical, i.e. non
(local) quantum case. Saying in an overstretched manner, the underlying
classical geometry of spacetime results in our approach from the quantum
theory by measuring the deviation from the commutative ''classical'' case.
(Notice that by the ''classical'' case we do not refer to any underlying
quantization procedure nor to a semiclassical approximation via some
perturbation theory. The route we follow in our reconstruction of spacetime
grounds in the noncommutativity and is intrinsically non pertubative.)

\section{\protect\medskip On the modular theory of double cones (free
massive case)}

Consider a double cone algebra $\mathcal{A}(\mathcal{O})$ generated by a
free massless field (for s=0 take the infrared convergent derivative). Then
the modular objects of ($\mathcal{A}(\mathcal{O}),\Omega )_{m=0}$ are
well-known \cite{Haa}. In particular the modular group is a one parametric
subgroup of the proper conformal group. The massive double cone algebra
together with the (wrong) massless vacuum has the same modular group $\sigma
_{t}$ however its action on smaller massive subalgebras inside the original
one is not describable in terms of the previous subgroup. In fact the
geometrical aspect of the action is wrecked by the breakdown of Huygens
principle, which leads to a nonlocal reshuffling inside $O$ but still is
local in the sense of keeping the inside and its causal complement apart.
This mechanism can be shown to lead to a pseudo-differential operator for
the infinitesimal generator of $\sigma _{t}$ whose's highest term still
agrees with conformal zero mass differential operator. We are however
interested in the modular group of ($\mathcal{A}(\mathcal{O}),\Omega )_{m}$
with the massive vacuum which is different from the that of the wrong vacuum
by a Connes cocycle. We believe that this modular cocycle will not wreck the
pseudo-differential nature, however we were not able to show this. We hope
that the above remarks may prove helpful in a future investigation.

\textbf{ACKNOWLEDGMENT} We would like to thank H.-J. Borchers and M. Schmidt
for several discussions and remarks.


\begin{thebibliography}{B/D/F/S}
\bibitem[Ar/Zs]{Ar/Zs}  H. Araki, L. Zsido ''An Extension of the Structure
Theorem of Borchers with an Application to Half-Sided Modular Inclusions''
preliminary version (1995)

\bibitem[Bi/Wi]{Bi/Wi}  J. Bisognano, E. Wichmann ''On the Duality Condition
for a Hermitean Scalar Field.'' Journ. Math. Physics \textbf{16} p. 985
(1975)

\bibitem[Bo1]{Bo1}  H.-J. Borchers ''The CPT-Theorem in Two-dimensional
Theories of Local Observables'' Comm. Math. Phys. \textbf{143 }p. 315 (1992)

\bibitem[Bo2]{Bo2}  H.-J. Borchers ''Half-Sided Modular Inclusions and the
Construction of the Poincar\'{e} Group '' Comm. Math. Physics \textbf{176 }%
p. 703 (1996

\bibitem[Bo3]{Bo3}  .-J. Borchers ''On Poincar\'{e} transformations and the
modular group of the algebra associated with a wedge'' preprint (1998), to
appear in Lett. Math. Phys.

\bibitem[Bo4]{Bo4}  H.-J. Borchers ''When Does Lorentz Invariance Imply
Wedge Duality'' Lett. Math. Phys. \textbf{35} p. 39 (1995)

\bibitem[Bo5]{Bo5}  H.-J. Borchers ''On the use of modular groups in quantum
field theory'' Ann. Henri Poincar\'{e} \textbf{63} p 331 (1995).

\bibitem[Bo/Yng]{Bo/Yng}  H.-J. Borchers, J. Yngvason ''Modular Groups of
Quantum Fields in Thermal States.'' preprint (1998)

\bibitem[Fr/Joe]{Fr/Joe}  K. Fredenhagen and Joerss, ''Conformal
Haag-Kastler nets, pointlike localized fields and the existence of operator
product expansions'', Comm. Math. Phys. \textbf{176}, (1996) p. 541

\bibitem[B/D/F/S]{B/D/F/S}  D. Buchholz, O. Dreyer, M. Florig and S.
Summers, ''Geometric Modular Action and Spacetime Symmetry Groups'' hep
9805026

\bibitem[G/L/W]{G/L/W}  D. Guido, R. Longo, H.-W. Wiesbrock ''Extensions of
Conformal Nets and Superselection Structures'' Comm. Math.Phys. \textbf{192 }%
p.217 (1998)

\bibitem[Haa]{Haa}  R. Haag ''Local Quantum Physics'' Springer Verlag (1992)

\bibitem[H/H/W]{H/H/W}  R. Haag, N. Hugenholtz, M. Winnink ''On the
equilibrium state in quantum statistical mechanics'' Comm. Math. Phys. 
\textbf{5 }p 215 (1967)

\bibitem[Ja]{Ja}  Ch. Jaekel ''Cluster Estimates for Modular Structure'' hep
9804017

\bibitem[Ka/Wi]{Ka/Wi}  R. Kaehler, H.-W. Wiesbrock ''Modular Theory and the
Reconstruction of 4-dim. Quantum field theories'' in preparation

\bibitem[Pe]{Pe}  G. Pedersen ''C*-algebras and their automorphism groups''
Academic Press, London, New York, San Francisco (1979)

\bibitem[O/L/R]{O/L/R}  P. Leyland, J. Roberts and D. Testard, ``Duality for
Quantum Fields'' unpublished preprint July 1978

J. P. Eckmann and K. Osterwalder, J. Funct. An. \textbf{13}, (1973) 1

\bibitem[Schr1]{Schr1}  B. Schroer, ''Motivations and Physical Aims of
algebraic Quantum Field Theory'' Amm. Phys. (N.Y.) \textbf{255}, (1997) 270.
The relation between the positive energy and the isotony of the net of wedge
subspaces of the Wigner representation one of the authors (B. S.) learned
from a talk by D. Guido 1996.

\bibitem[Schr2]{Schr2}  B. Schroer, ''Modular Wedge Localization and the
d=1+1 Formfactor Program'' hep-th/9712124, to be published in AOP.

\bibitem[Schr3]{Schr3}  B. Schroer, a course on ''Localization and
Nonperturbative Local Quantum Physics'', hep-th/9805093, chapter 3 and 6

\bibitem[Se]{Se}  G. Sewell ''Quantum Fields on Manifolds: PCT and
Gravitationally Induced Thermal States'' Ann. of Phys. \textbf{141 }p 201
(1982)

\bibitem[Stra]{Stra}  S. Stratila ''Modular Theory in Operator Algebras''
Abacus Press 1981.

\bibitem[Ta]{Ta}  M. Takesaki ''Tomita's Theory of Modular Hilbert Algebras
and its Applications'' Lect. Notes in Math. \textbf{128 }Springer Verlag
(1970)

\bibitem[Un]{Un}  W. Unruh ''Notes on Black Hole Evaporation'' Phys. Rev. 
\textbf{D29 }p 1047 (1976)

\bibitem[Was]{Was}  A. Wassermann, ''Operator Algebras and Conformal Field
Theory'', University of Cambridge preprint 1997

\bibitem[Wi1]{Wi1}  H.-W. Wiesbrock ''Half-Sided Modular Inclusions of
von-Neumann Algebras'' Comm. Math. Phys. \textbf{157 }p 83 (1993) \textbf{%
Erratum }Comm. Math. Phys. \textbf{184 }p.683 (1997)

\bibitem[Wi2]{Wi2}  H.-W. Wiesbrock ''Conformal Quantum Field Theory and
Half-Sided Modular Inclusions of von-Neumann Algebras.'' Comm. Math. Phys. 
\textbf{158 }p 537 (1993)

\bibitem[Wi3]{Wi3}  H.-W. Wiesbrock ''Symmetries and Modular Intersections
of von-Neumann Algebras'' Lett. Math. Phys. \textbf{39 }p 203 (1997)

\bibitem[Wi4]{Wi4}  H.-W. Wiesbrock ''Modular Intersections of von-Neumann
Algebras in Quantum Field Theory'' Comm. Math. Phys. \textbf{193 }p 269
(1998)\textbf{\ }

\bibitem[Yng]{Yng}  J. Yngvason ''A Note on Essential Duality'' Lett. Math.
Phys. \textbf{31 }p 127 (1995)
\end{thebibliography}
\end{document}